\documentclass[a4paper,12pt]{article}
\usepackage[english]{babel}
\usepackage[utf8]{inputenc}
\usepackage{url}
\usepackage{layaureo}
\usepackage[dvips]{epsfig}

\usepackage{graphicx}
\usepackage{color}
\usepackage{multirow}

\usepackage{amsthm}
\usepackage{amsmath}
\usepackage{amsfonts}
\usepackage{amssymb}
\usepackage{graphicx}
\usepackage{setspace}
\usepackage{natbib}
\usepackage{booktabs}
\usepackage[colorlinks,breaklinks]{hyperref}
\hypersetup{
    colorlinks,%
    citecolor=black,%
    filecolor=black,%
    linkcolor=black,%
    urlcolor=black
}
\usepackage{lscape}
\usepackage{ctable}

\usepackage{array}
\newcolumntype{C}[1]{>{\centering\let\newline\\\arraybackslash\hspace{0pt}}m{#1}}

\usepackage{pdflscape}
\usepackage{longtable}
\numberwithin{equation}{section}

\begin{document}

\begin{center}
{\noindent \textbf{\Large Text
mining arXiv: a look through \\ quantitative finance papers}}

\vspace{14pt}

{\noindent Michele Leonardo Bianchi\textsuperscript{a,}\footnote{ The author thanks arXiv, ChatGPT, and Google Scholar for use of their open access interoperability and Sabina Marchetti for her comments and suggestions. This publication should not be reported as representing the views of the Bank of Italy. The views expressed are those of the author and do not necessarily reflect those of the Bank of Italy.}}

\vspace{14pt}

{\noindent \small\textsuperscript{a}\it Financial Stability Directorate, Bank of Italy, Rome, Italy}\\
%

\vspace{20pt}

This version: \today

\end{center}

\vspace{14pt}

\noindent {\bf Abstract.} This paper explores articles hosted on the arXiv preprint server with the aim to uncover valuable insights hidden in this vast collection of research. Employing text mining techniques and through the application of natural language processing methods, we examine the contents of quantitative finance papers posted in arXiv from 1997 to 2022. We extract and analyze crucial information from the entire documents, including the references, to understand the topics trends over time and to find out the most cited researchers and journals on this domain. Additionally, we compare numerous algorithms to perform topic modeling, including state-of-the-art approaches. 

\vskip 0.6cm


\section{Introduction}\label{sec:Introduction}

Quantitative finance is a field of finance that studies mathematical and statistical models and applies them to financial markets and investments, for pricing, risk management, and portfolio allocation. These models are needed to analyze financial data, to find the price of financial instruments and to measure their risk (see \cite{vogl2022quantitative} and \cite{btf2023fat}). Readers are referred to \cite{derman2011models} for an insightful exploration of the role of models in finance and to \cite{ippoliti2021mathematics} for some philosophical remarks on mathematics and finance. 

The world of finance is always moving forward even in times of crisis. Innovations in finance come from the development of new financial services, products or technologies. Research trends in quantitative finance are driven not only by innovations, but also by structural changes in financial markets or by changes in regulation (\cite{cesa2017brief} and \cite{carmona2022influence}). When a structural change occurs some models are not anymore able to explain the phenomena observed in the market, consequently quants and researchers start working on new models. Examples of such changes are when the implied volatility smile appeared in 1987 (see \cite{derman2016volatility}) or the Euribor-OIS spread materialized in 2007 (see \cite{bianchetti2008two}). Research activities driven by new products are, for instance, the development of pricing models for interest rate and equity derivatives started in '90s, the structuring of credit products in the early 2000s, or the recent research trend on cryptocurrencies. New technologies applied to finance are namely the increasing role of big data and the advent of machine learning techniques. Regulation have an impact on the development of new quantitative tools for measuring, managing and monitoring financial risks (e.g. the Basel Accords). 

In this paper we explore the arXiv preprint server, the dominant open-access preprint repository for scholarly papers in the fields of physics, mathematics, computer science, quantitative biology, quantitative finance, statistics, electrical engineering and systems science, and economics. The papers in arXiv are not peer-reviewed but there are advantages in submitting to this repository, mainly to disseminate a paper without waiting for the peer review and publishing process, which can be slow (see \cite{huisman2017duration}). The arXiv collection provides a unique source of data to conduct various studies, including bibliometric, trend, and citation network analyses (see \cite{clement2019use}). It is a valuable resource for advancing scientific knowledge and conducting research on research, often referred to as {\it meta-research}. For example, \cite{eger2019predicting} and \cite{viet2019trends} perform trend detection on computer science papers stored in arXiv, \cite{lin2020many} conduct a case study of computer science preprints submitted to arXiv from 2008 to 2017 to quantify how many preprints have eventually been printed in peer-reviewed venues, \cite{tan2021images} explore the images of around 1.5 million of papers held in the repository, \cite{okamura2022scientometric} investigates the citations of more than 1.5 million preprints on arXiv to study the evolution of collective attention on scientific knowledge, \cite{bohara2023roberta} train a state-of-art classification approach, and \cite{fatima2023retrieving} design an algorithm to help researchers to perform systematic literature reviews. 

We study all papers on quantitative finance, a small portion of the entire arXiv containing more than two millions of works at the time of writing. The choice is also motivated by our experience in this domain and by scientific curiosity. 

The code is run on a standard desktop environment, without recurring to a big cluster. Scaling to a large number of papers may be not trivial. Dealing with a large amount of data requires significant computing resources, including processing power and memory, to manipulate and analyze the data efficiently. It is not simple, and maybe even impossible, to explore more than two million of papers with a standard desktop environment like ours.

The studies of papers on finance topics is not new in the literature. \cite{burton2020twenty} review the history of a well-known journal in this field and highlight its growth in terms of productivity and impact. The authors present a bibliometric analysis and identify key contributors, themes, and co-authorship patterns and suggest future research directions. \cite{ali2022bibliometric} conduct a systematic literature review and a bibliometric analysis on around 3,000 articles on asset pricing sourced from the top 50 finance and economics journals, spanning a 47-year period from 1973 to 2020. As observed by the authors, the exclusion of certain publications may potentially offer an alternative perspective on the landscape of existing asset pricing research. By using bibliometric and network analysis techniques, including the Bibliometrix Tool of \cite{aria2017bibliometrix},  \cite{sharma2023review} investigate   more than 4,000 papers on option pricing appeared from 1973 to 2019. They follow the procedure suggested by \cite{donthu2021conduct}. Their study aims to pinpoint high-quality research publications, to discern trends in research, to evaluate the contributions of prominent researchers, to assess contributions from different geographic regions and institutions, and, ultimately, to examine the interconnectedness among these aspects. The works of \cite{burton2020twenty}, \cite{ali2022bibliometric} and \cite{sharma2023review} are focused on asset pricing or on a specific journal, their corpus is obtained by searching in the Scopus database some specific keywords, and the bibliometric analysis relies on VOS viewer (see \cite{van2010software}) and Gephi (see \cite{bastian2009gephi}). 

Our study explores all papers on quantitative finance collected in arXiv till the end of 2022 (around 16,000) and it considers text mining techniques implemented in Python to extract information directly from the portable document format (pdf) files containing the full text of the papers, excluding images, without relying on ad-hoc software or proprietary databases. \cite{westergaard2018comprehensive} found that examining the full text of documents significantly improved text mining compared to studies that only explored information collected from abstracts.\footnote{ As a crosscheck, we conduct the analysis on both abstracts and full texts. The analysis using the full text data shows better results.} Their finding highlights the importance of using complete textual content for more comprehensive and accurate text mining and analysis.

The main objectives of our work are twofold. First, we explore the topics of the quantitative finance papers collected in arXiv in order to describe the evolution of topics over time. After having evaluated the performance of various clustering algorithms, we investigate on which themes researchers have focused their attention in the period from 1997 to 2022. Second, we try to understand who are the most prominent authors and journals in this field. Both analyses are performed with data mining techniques and without actually reading the papers. 

The remainder of the paper is organized as follows. First, we provide a brief description of the data analyzed in this work (Section \ref{sec:Data}). Then, in Section \ref{sec:Preprocessing} the preprocessing phase is discussed by offering further insights on the papers analyzed in our work. In Section \ref{sec:Topics} we compare various clustering algorithms and, after having selected the best performer, we explore, by splitting our corpus in 30 clusters, the evolution of topics over time. Finally, in Section \ref{sec:Authors} we describe an entity extraction process to investigate authors and journals with the largest number of occurrences in the corpus considered in this work.  Section \ref{sec:Conclusions} concludes.

\section{Data description}\label{sec:Data}

In this section we provide a description of the papers analyzed in this work. As observed above, there are various domains in arXiv (i.e. physics, mathematics, computer science, quantitative biology, quantitative finance, statistics, electrical engineering and systems science, and economics) and each domain has is own categories. The categories within the quantitative finance domain are the following:

\begin{itemize}
\item computational finance (q-fin.CP) includes Monte Carlo, PDE, lattice and other numerical methods with applications to financial modeling;
\item economics (q-fin.EC) is an alias for econ.GN and it analyses micro and macro economics, international economics, theory of the firm, labor economics, and other economic topics outside finance;
\item general finance (q-fin.GN) is focused on the development of general quantitative methodologies with applications in finance;
\item mathematical finance (q-fin.MF) examines mathematical and analytical methods of finance, including stochastic, probabilistic and functional analysis, algebraic, geometric and other methods;
\item portfolio management (q-fin.PM) deals with security selection and optimization, capital allocation, investment strategies and performance measurement;
\item pricing of securities (q-fin.PR) discusses valuation and hedging of financial securities, their derivatives, and structured products;
\item risk management (q-fin.RM) is about risk measurement and management of financial risks in trading, banking, insurance, corporate and other applications;
\item statistical finance (q-fin.ST) includes statistical, econometric and econophysics analyses with applications to financial markets and economic data;
\item trading and market microstructure (q-fin.TR) studies market microstructure, liquidity, exchange and auction design, automated trading, agent-based modeling and market-making.
\end{itemize}

\begin{figure}
\begin{center}
\includegraphics[width=\columnwidth]{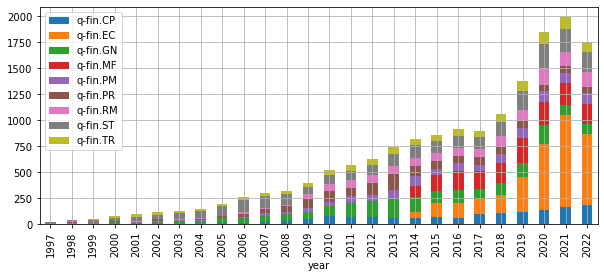}
\caption[Category]{\label{fig:Categories}\footnotesize Categories by year.}
\end{center}
\end{figure}

These categories are assigned by the authors when they submit their papers. Even if it is possible to select multiple couples of domain-category belonging to more than one domain, we select as reference category only the first category within the quantitative finance domain. Figure \ref{fig:Categories} shows the numbers of papers on quantitative finance submitted to arXiv between 1997 and 2022. The increase in the last three years is mainly due to the q-fin.EC category.

As observed in Section \ref{sec:Introduction}, the code is implemented in Python and it is run under Ubuntu 22.04 on a desktop with an AMD Ryzen 5 5600g processor with 32GB of RAM. As we will describe in the following, numerous packages are considered.

As far as the collection process is concerned, we retrieve data from arXiv by selecting all categories within quantitative finance (i.e. q-fin). We collect articles metadata and pdf files for all articles from 1997 to 2022 for a total of around 16,000 articles (18GB of data). 

While the metadata are obtained through {\it urllib.request} and {\it feedparser}, the pdf files are downloaded by means of the {\it arxiv} package. The metadata can be collected by following the suggestions provided in the arXiv web-pages. They are a fundamental input of the analysis and include the link to the paper main web-page, from where it is possible to extract the paper identification code ({\it id}, e.g. 2005.06390). The metadata contain information like authors names,  paper title, primary category, submission and last update dates, abstract, and publication data when available (e.g. digital object identifier, DOI). Subsequent updates of the papers can be stored in the repository and for this reason there is the version number at the end of the paper {\it id}. 

Since a paper could be assigned to multiple categories, a web-scraping tool written in Python allows us to retrieve from the paper main web-page the list of all categories of each single paper. We select from this list the subset of categories within the quantitative finance domain. Thus we assign as  reference category of a paper the first category appearing in this subset. Starting from this list, we are able to filter and analyze all papers in the nine categories within q-fin. 

\begin{figure}
\begin{center}
\includegraphics[width=\columnwidth]{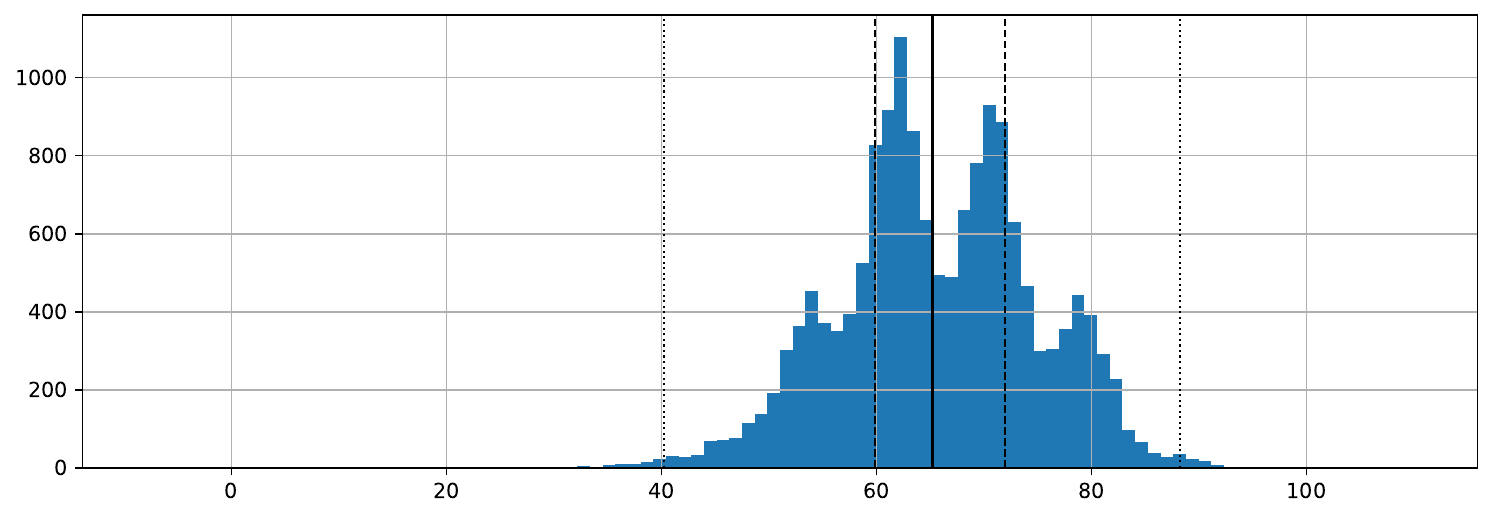}
\caption[ ]{\footnotesize The Flesch reading ease score is reported. The vertical line represents the median, the dashed line is the quantile of level 0.25 (0.75), and the dotted line is the quantile of level 0.005 (0.995).}
\label{fig:Readability}
\end{center}
\end{figure}

The {\it pdftotext} package allows us to extract the text from pdf files. Each paper becomes a single (long) string. As discussed in Section \ref{sec:Preprocessing}, the length of these strings vary accross papers (see Figure \ref{fig:Length}), also because some documents are not papers (e.g. there are both theses and books). As a first assessment of the corpus, for each document we estimate the readability of the papers through {\it textstat}. As shown in Figure \ref{fig:Readability} the Flesch reading ease score (see \cite{dubay2004principles}) is on average equal to 65.7 (plain English), the lower and upper quartile are 59.91 (fairly difficult to read, but not far from the plain English) and 71.95 (fairly easy to read), respectively, and 
99 per cent of the papers are in the range from 40.28 (difficult to read) and 88.20 (easy to read). There is only one paper with a negative value, but this is caused by the text contained in the figures. All other papers are above 17.17, that is above the {\it extremely difficult to read} level.

\section{Text preprocessing}\label{sec:Preprocessing}

This section describes the text processing steps. The preprocessing phase is performed with {\it nltk}:(1) we split the text in tokens; (2) we extract the numbers representing years in the text;\footnote{ We assume these numbers have 4 digits. We do not explore this data in the empirical analysis.} (3) we identify all strings containing alphabet letters, and we refer to them as {\it words} even in the case they do not belong to the English vocabulary; this step allows also to remove some symbols which are not recognized as letters in text analysis. Then, (4) we remove all stopwords and all words with length less than 3 characters; we also check whether there are words with more than 25 characters (quite uncommon in English); (5) we conduct a lemmatization by means of a part-of-speech tagger considering nouns, verbs, adjectives and adverbs; (6) we check if the paper is written in English by means of the {\it langdetect} package and we discard all non English papers. Both the extraction phase and the preliminary text analysis is parallelized by means of the {\it multiprocessing} package. We refer to the output of this first preprocessing phase as \textit{lemmatized data}.

\begin{figure}
\begin{center}
\includegraphics[width=1.5\columnwidth,angle=-90]{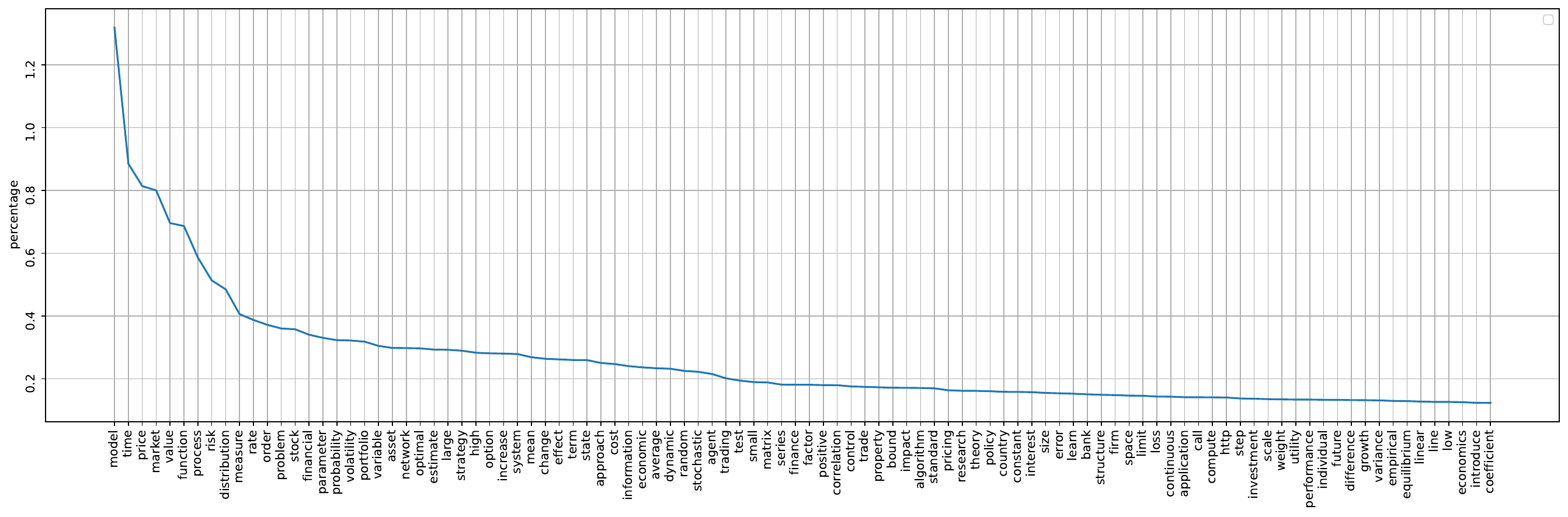}
\caption[ ]{\label{fig:FrequentWords}\footnotesize Frequent words of the corpus and their percentage of appearance.}
\end{center}
\end{figure}

\begin{figure}
\begin{center}
\includegraphics[width=1\columnwidth]{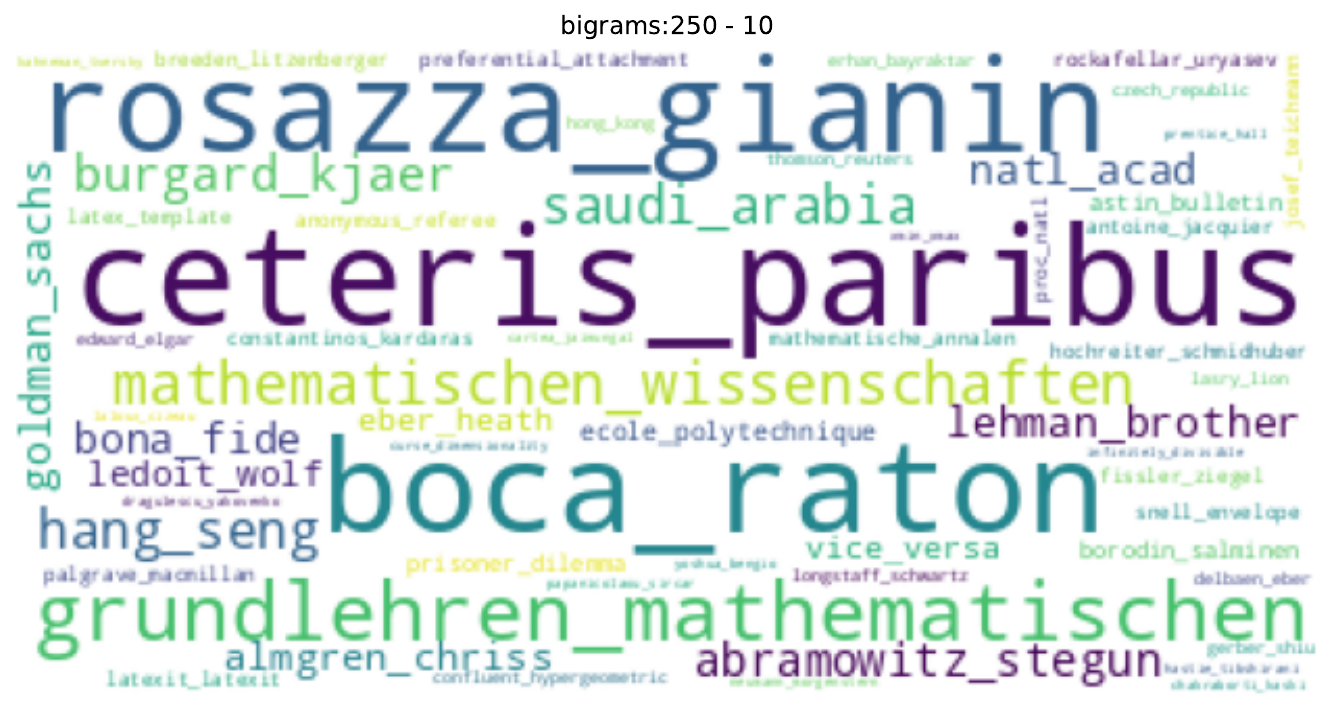}
\includegraphics[width=1\columnwidth]{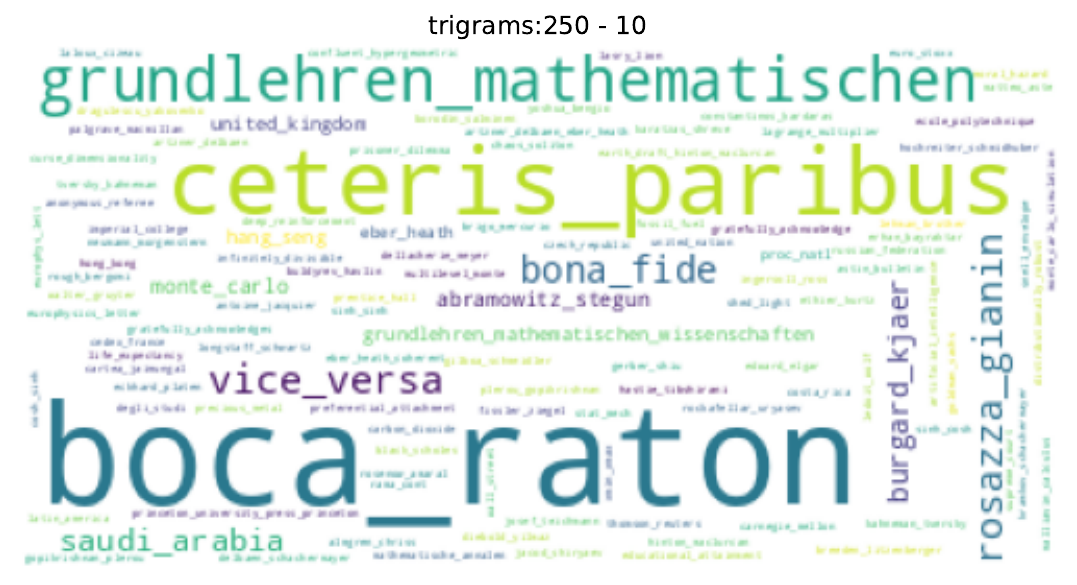}
\caption[ ]{\label{fig:WordCloud}\footnotesize Bigrams and tri(four)grams word clouds based on frequency with parameters \textit{min count} equal to 250 and \textit{threshold} equal to 10.}
\end{center}
\end{figure}

Thus, we analyze the frequency of the words across the whole corpus and we remove all words appearing less than 25 times. We discard also some words frequently used in writing papers on quantitative finance and which do not help in understanding the topic of the paper. The list of these words includes, for example, ``proof'' and ``theorem'', verbs commonly used in mathematical sentences (e.g.,``assume'',``satisfy'', and ``define''), mathematical functions (e.g. ``min'', and ``log'') and adverbs. The complete list is available upon request. In Figure \ref{fig:FrequentWords} the list of the top 100 most frequent words obtained after this cleaning phase and their percentage of appearance is shown. The word ``model'' is extremely frequent (one every 100 words).
The word ``http'' is also quite common, indicating that the papers full texts contain numerous internet links.

After a first preprocessing phase, we conduct an $n$-gram analysis by considering the {\it Phrases} model of the {\it gensim} package. We ignore all words and bigrams with total collected count over the entire corpus lower than 250 and set the score threshold equal to 10. We find the bigrams and then, to find trigrams and fourgrams, we apply again the same model to the transformed corpus including bigrams. This approach allows us to have a better ex-post understanding of the corpus which is full of $n$-grams (e.g. Monte Carlo simulation,  Eisenberg and Noe, or bank balance sheet). The wordclouds of the main bigrams and trigrams (fourgrams) are depicted in Figure \ref{fig:WordCloud}.

It should be noted that some topic modeling algorithms analyzed in Section \ref{sec:Topics} do not need text preprocessing. In those cases the input is just a single list containing the whole paper text. While we refer to this latter input as \textit{raw data}, we define the output of the preprocessing as \textit{cleaned data}. 


\begin{figure}
\begin{center}
\includegraphics[width=\columnwidth]{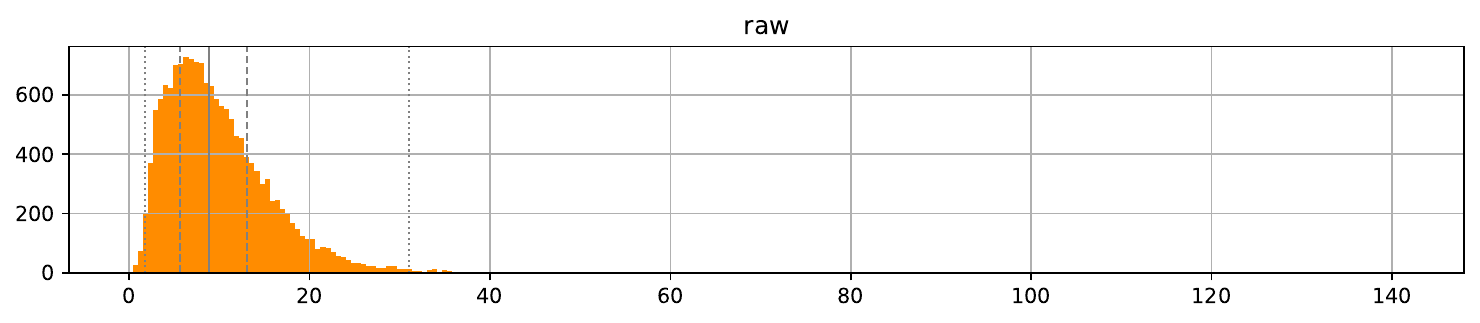}
\includegraphics[width=\columnwidth]{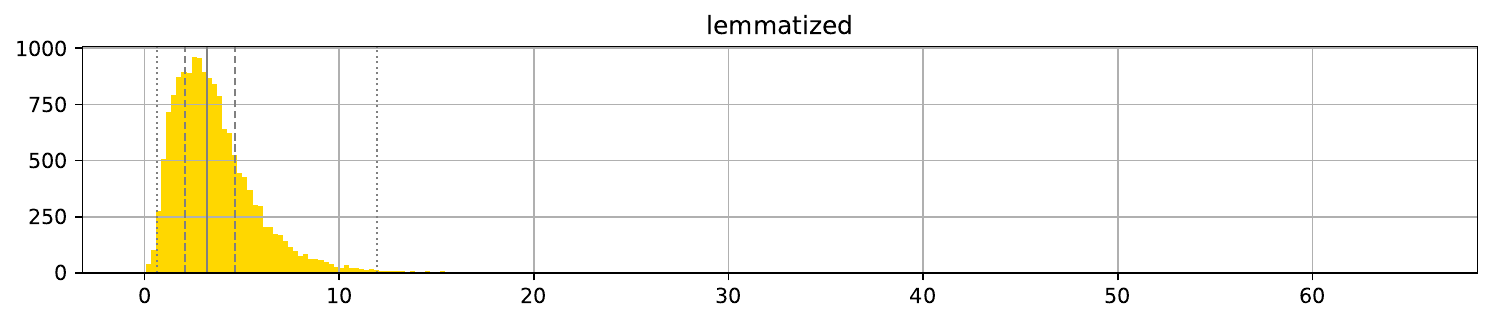}
\includegraphics[width=\columnwidth]{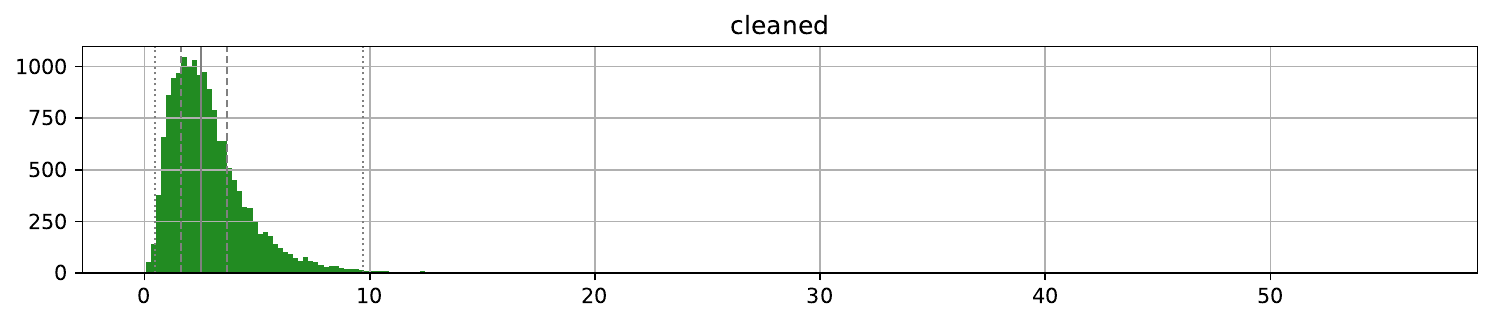}
\caption[ ]{\footnotesize Papers length, in terms of number of words, for raw, lemmatized and cleaned data. The x-axis values are in thousands. The scale of x-axis varies across the three datasets. The vertical line represents the median, the dashed line is the quantile of level 0.25 (0.75), and the dotted line is the quantile of level 0.01 (0.99).}
\label{fig:Length}
\end{center}
\end{figure}

In Figure \ref{fig:Length} we show how the length of the papers varies over the three data preprocessing phases. Starting from the raw data, containing all words and symbols, after a preliminary cleaning step we obtain the lemmatized data and then, after the last cleaning steps, the cleaned data. The number of words varies from a median value of 8,824 words for the raw data to around 2,518 words for the cleaned ones.  

\section{Topics trend}\label{sec:Topics}

Now we are in the position to perform a topics trend analysis. We employ a topic modeling approach to identify the subjects discussed in the documents examined in this study, and then we observe how these topics change over time. Topic modeling refers to a class of statistical methods used to determine which subjects are prevalent in a given corpus. In Section \ref{sec:AlgorithmPerformance} we select the best performing model among some selected approaches presented in the related literature. We evaluate these approaches by assessing their ability to accurately match the nine q-fin categories that researchers assign to their work when submitting it to arXiv. Then in Section \ref{sec:Study}, after having split the papers into 30 clusters, each one representing a specific topic, we discuss the evolution of research trends over time.

\subsection{Algorithms performance on full texts}\label{sec:AlgorithmPerformance}

Topic modeling algorithms are widely used in natural language processing and text mining to uncover latent thematic structures in a collection of documents. Different algorithms have been developed, each with its own strengths and limitations (see \cite{sethia2022framework}). The choice of a topic modeling algorithm depends on specific factors, such as the desired level of topic granularity and computational constraints. Some algorithms may require substantial computational resources and large amounts of training data. Here, we compare different algorithms by looking at some standard performance measures. 

Since is not simple to assess the performance of different topic modeling algorithms (see \cite{rudiger2022topic}), we start by comparing the clusters assigned by each algorithm on the entire corpus to the nine clusters defined by the q-fin categories described in Section \ref{sec:Data}, that is the categories that researchers assign to their work during the submission process to arXiv. By exploiting a Bayesian optimization strategy, \cite{terragni2021octis} present a framework for training, analyzing, and comparing topic models where the competitor models are trained by searching for their optimal hyperparameter configuration, for a given metric and dataset. Here we consider a simpler approach in which we compare the models by looking at some metrics. Evaluating topic modeling algorithms typically involves the use of performance measures, and it is important to note that different algorithms can yield varying results across these metrics.

We consider the following models.

\begin{itemize}
\item \textit{K-means}. We perform a clustering analysis by considering the K-means algorithm implemented in \textit{scikit-learn}. This algorithm groups data points into K clusters by minimizing the distance between data points and their cluster center. The document word matrix is created through the \textit{CountVectorizer} function, that converts the corpus to a matrix of token counts. We ignore terms that have a document frequency strictly higher than 75\%. T
\item \textit{LDA}. By considering the same document word matrix analyzed with the K-means algorithm, we perform topic modeling with the latent Dirichlet allocation (LDA). LDA is a well-known unsupervised learning algorithm. As observed in the seminal work of \cite{blei2003latent}, the basic idea is that documents are represented as random mixtures over latent topics, where each topic is characterized by a distribution over words. We study two different implementations of LDA (i.e. \textit{scikit-learn} and \textit{gensim}). 
\item \textit{Word2Vec}. We train a word embedding model (i.e. Word2Vec) and then we perform a clustering analysis by considering again the K-means approach. An embedding is a low-dimensional space into which high-dimensional vectors are projected. Machine learning on large inputs like sparse vectors representing words is easiser if embeddings are considered. Ideally, an embedding captures some of the semantics of the input by placing semantically similar inputs close together in the embedding space. The Word2Vec neural network introduces distributed word representations that capture syntactic and semantic word relationships (see \cite{mikolov2013efficient}). More in details, we generate document vectors using the trained Word2Vec model, that is, we  get numerical vectors for each word in a document, and then the document vector is the weighted average of the vectors. Thus, the K-means algorithm is applied to the matrix representing the corpus. We consider the Word2Vec model implemented in \textit{gensim}.
\item \textit{Doc2Vec}. We create a vectorised representation of each document through the Doc2Vec model and then we perform a clustering analysis by considering the K-means approach. The Doc2Vec extends Word2Vec and it can learn distributed representations of varying lengths of text, from sentences to documents (see \cite{le2014distributed}). We consider the Doc2Vec model implemented in \textit{gensim}.
\item \textit{Top2Vec}. We study the Top2Vec model, an unsupervised learning algorithm that finds topic vectors in a space of jointly
embedded document and word vectors (see \cite{angelov2020top2vec}). This algorithm directly detects topics by performing the following steps. First, embedding vectors for documents and words are generated. Second, a dimensionality reduction on the vectors is implemented. Third, the vectors are clustered and topics are assigned. This algorithm is implemented in an ad-hoc library named \textit{Top2Vec} and it automatically provides information on the number of topics, topic size, and words
representing the topics.
\item \textit{BERTopic.} We study a BERTopic model, which is similar to Top2Vec in terms of algorithmic structure and uses BERT as an embedder. As described in the seminal work of \cite{grootendorst2022bertopic}, from the clusters of documents, topic representations are extracted using a custom class-based variation of term frequency-inverse document frequency (TF-IDF). This is the main difference with respect to Top2Vec. The algorithm is implemented in an ad-hoc library named \textit{BERTopic}. The main downside with working with large documents, as in our case, is that information will be ignored if the documents are too long.  A limited number of tokens are treated and anything longer is cut off. Since we are dealing with large documents, to work around this issue, we first split each documents in chunks of 300 tokens, thus we fit the model on these chunks.
BERTopic does not allow one to directly select the number of topics, for this reason on the first step we obtain a number of topics much larger than the desired one. 
Since we obtain for each chunk the corresponding topic, we have for each document a list of possibly different topics and the length of these lists varies across documents (i.e. the length of a single list depends on the length of the corresponding document). To cluster this list of lists of topics, we consider each integer representing a topic as a word. Thus we use the Word2Vec algorithm described above to find similarities between these list of topics. Each topic label, that is the number representing the topic, is treated as a string, and Word2Vec transforms it into a numerical vector. We then apply K-means clustering to group these lists based on their similarity in the vector space. The resulting clusters reveal relationships and patterns among these lists and allow us to select the number of clusters we need for our purposes.
\end{itemize}

In theory both BERTopic and Top2Vec should use raw data since these algorithms rely on an embedding approach, and keeping the original structure of the text is of paramount importance (see \cite{egger2022topic}). However, raw data extracted from quantitative finance papers have a considerable amounts of formulas, symbols and numbers that may affect the algorithms performance. For this reason, we consider both raw and cleaned data. Additionally, for these state-of-the-art algorithms the number of extracted topics tends be large, but the algorithms offer the possibility to reduce the number of topics and of outliers, that can be larger than expected. The parameters of a BERTopic model have to be carefully chosen to avoid memory issues. Alternatively, it is possible to perform an online topic modeling, that is the model is trained incrementally from a mini-batch of instances. This result in a less resource demanding approach in terms of memory and CPU usage, but it generates less rich and less comprehensive outputs and for these reasons we do not consider this incremental approach here.

In Table \ref{table:sumStatLog} we report the following similarity measures between true and predicted cluster labels: 
(1) the rand score (RS) is defined as the ratio between the number of agreeing pairs and the number of pairs, and it ranges between 0 and 1, where 1 stands for perfect match; (2) the adjusted rand score (ARS), that is the rand score adjusted for chances, has a value close to 0 for random labeling, independently of the number of clusters and samples, and exactly 1 when the clusterings are identical (up to a permutation), however is bounded below by -0.5 for especially discordant clusterings; (3) the mutual info score (MI) is independent of the absolute values of the labels (i.e. a permutation of the cluster labels does not change the value of the score); 
(4) the normalized mutual information (NMI) is a normalization of the MI to scale the results between 0 (no mutual information) and 1 (perfect correlation); 
(5) cluster accuracy (CA) is based on the Hungarian algorithm to find the optimal matching between true and predicted cluster labels; (6) to compute the purity score (PS), each cluster is assigned to the class which is most frequent in the cluster, and the similarity measure is obtained by counting the number of correctly assigned papers and dividing by the number of observations. This latter score increases as the number of clusters increases and for this reason, it cannot be used as a trade off between the number of clusters and clustering quality, that is to find the optimal number of clusters.

The measures presented in Table \ref{table:sumStatLog} demonstrates that the Doc2Vec approach, when coupled with K-means clustering on cleaned data, outperforms other models. This is evident from the higher MI and PS measures it achieves compared to its competitors. Moreover, Doc2Vec exhibits practical advantages, as it is straightforward to implement and significantly reduces computing time when compared to more advanced techniques like BERTopic. Interpreting the results of the Doc2Vec approach is simple, as it allows for the easy identification of the most representative documents by retrieving the centroid vectors of each cluster. The Word2Vec approach, when coupled with K-means clustering on cleaned data, shows also a good performance.

\begin{table}
\centering
\begin{footnotesize}
\begin{tabular}{lC{1.5cm}C{1.5cm}C{1.5cm}C{1.5cm}C{1.5cm}C{1.5cm}}
  \hline
	&	RS & ARS  &  MI & NMI & CA & PS \\
  \hline
K-means & 0.570 & 0.029 & 0.232 & 0.136 & 0.271 & 0.297 \\
LDA scikit-learn & 0.823 & 0.194 & 0.608 & 0.284 & 0.376 & 0.460 \\
LDA gensim & 0.788 & 0.085 & 0.276 & 0.131 & 0.275 & 0.314  \\
Word2Vec K-means & {\bf 0.832} & 0.200 & 0.613 & 0.283 & 0.371 & 0.427 \\
Doc2Vec K-means & 0.831 & 0.220 & {\bf 0.699} & 0.325 & 0.388 & {\bf 0.490} \\
Top2Vec raw & 0.810 & 0.195 & 0.501 & 0.239 & 0.365 & 0.404 \\
Top2Vec cleaned & 0.811 & 0.206 & 0.530 & {\bf 0.387} & 0.253 & 0.416 \\
BERTopic raw & 0.826 & 0.238 & 0.608 & 0.289 & {\bf 0.436} & 0.458 \\
BERTopic cleaned & 0.821 & {\bf 0.239} & 0.574 & 0.276 & 0.398 & 0.429 \\
 \hline
\end{tabular}
\caption{\footnotesize Algorithms performance. We report the rand score (RS), the adjusted rand score (ARS), the mutual info score (MI), the normalized mutual info score (NIM), the cluster accuracy (CA), and the purity score (PS).}
\label{table:sumStatLog} 
\end{footnotesize}
\end{table}

It is worth noting that there are no significant differences in performance measures when applying either the Top2Vec or BERTopic methods to raw or cleaned data. This could be attributed to the fact that raw data contain mathematical formulas that do not contribute substantial additional information, even if, as shown in Figure \ref{fig:Length}, raw data have a larger number of words. This finding indicating an equivalence between raw or cleaned data could be also attributed to the relatively simple structure commonly found in quantitative finance papers. It is important to note that these findings may not generalize to papers or books with more intricate and complex text structures and without formulas.

LDA implemented in {\it scikit-learn} has a better performance than the LDA implementation in {\it gensim}. The plain K-means does not show satisfactory results, even if the algorithm can be implement without a great effort.

Finally, as an overall assessment, it is important to highlight that the performance metrics reported in Table \ref{table:sumStatLog} do not demonstrate particularly impressive results. This could partly stem from the wide-ranging nature of each q-fin category, encompassing numerous subtopics and arguments. Conversely, some papers can be classified under multiple categories, as it is not always obvious how to select a single definitive category for a given work.

\subsection{Empirical study}\label{sec:Study}

\begin{figure}
\begin{center}
\includegraphics[width=0.19\columnwidth]{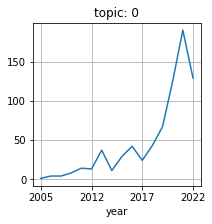}
\includegraphics[width=0.19\columnwidth]{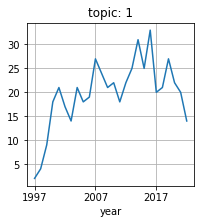}
\includegraphics[width=0.19\columnwidth]{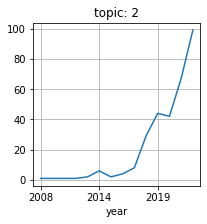}
\includegraphics[width=0.19\columnwidth]{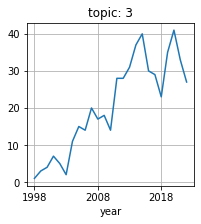}
\includegraphics[width=0.19\columnwidth]{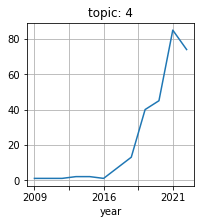}
\includegraphics[width=0.19\columnwidth]{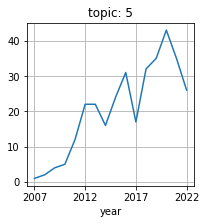}
\includegraphics[width=0.19\columnwidth]{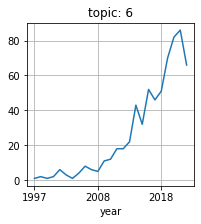}
\includegraphics[width=0.19\columnwidth]{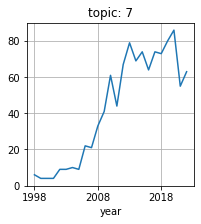}
\includegraphics[width=0.19\columnwidth]{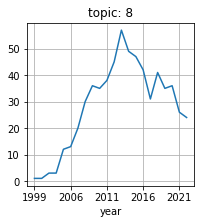}
\includegraphics[width=0.19\columnwidth]{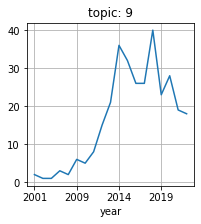}
\includegraphics[width=0.19\columnwidth]{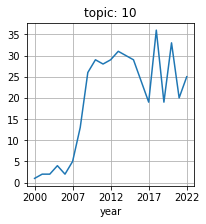}
\includegraphics[width=0.19\columnwidth]{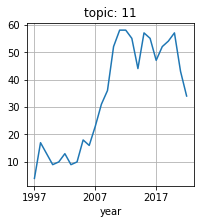}
\includegraphics[width=0.19\columnwidth]{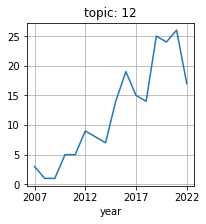}
\includegraphics[width=0.19\columnwidth]{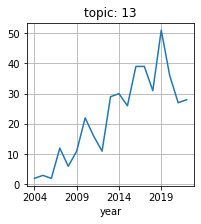}
\includegraphics[width=0.19\columnwidth]{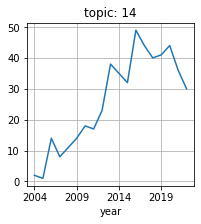}
\includegraphics[width=0.19\columnwidth]{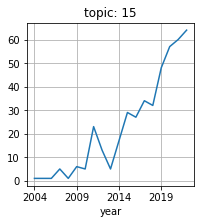}
\includegraphics[width=0.19\columnwidth]{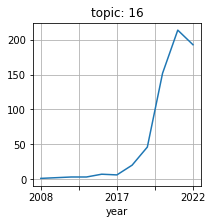}
\includegraphics[width=0.19\columnwidth]{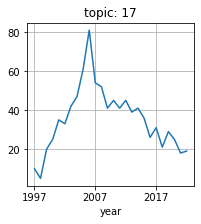}
\includegraphics[width=0.19\columnwidth]{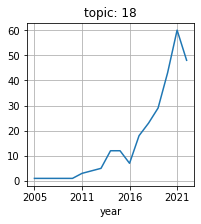}
\includegraphics[width=0.19\columnwidth]{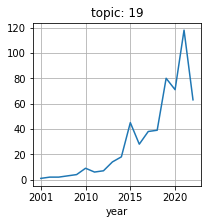}
\includegraphics[width=0.19\columnwidth]{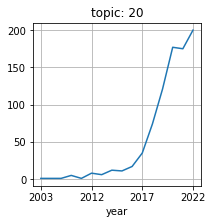}
\includegraphics[width=0.19\columnwidth]{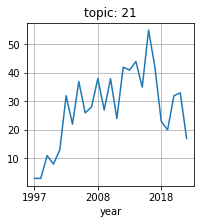}
\includegraphics[width=0.19\columnwidth]{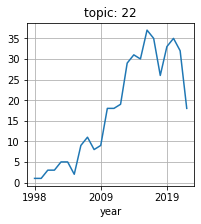}
\includegraphics[width=0.19\columnwidth]{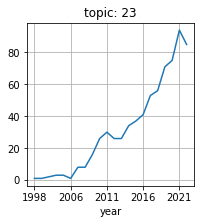}
\includegraphics[width=0.19\columnwidth]{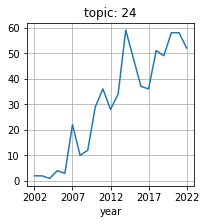}
\includegraphics[width=0.19\columnwidth]{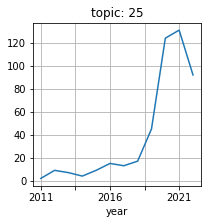}
\includegraphics[width=0.19\columnwidth]{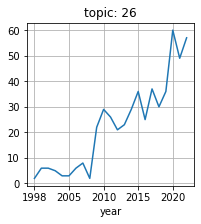}
\includegraphics[width=0.19\columnwidth]{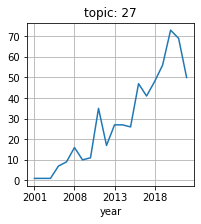}
\includegraphics[width=0.19\columnwidth]{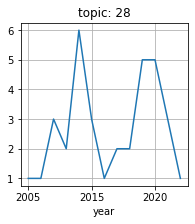}
\includegraphics[width=0.19\columnwidth]{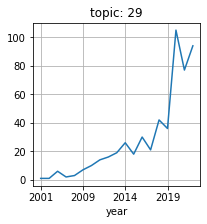}
\caption[Topics]{\label{fig:Topics}\footnotesize Topics trend by year across the sample of around 16,000 papers in the q-fin categories.}
\end{center}
\end{figure}

\begin{figure}
\begin{center}
\includegraphics[width=0.6\columnwidth]{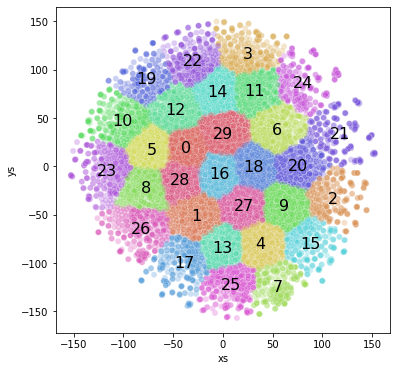}
\caption[Topics clusters]{\label{fig:TopicsClusters}\footnotesize We show the topics clusters projected on the 2-dimensional space through the standard t-distributed stochastic neighbor embedding (t-SNE) algorithm. The numbers represent the topics reported in Table \ref{table:Topics}.}
\end{center}
\end{figure}

As shown in Section \ref{sec:AlgorithmPerformance}, the best performing model is Doc2Vec with K-means clustering applied on cleaned data. This model is considered to have a better understanding of the topics discussed in the quantitative finance papers analyzed in this work. To obtain the desired number of topics we perform again a K-means clustering analysis. To extract the most representative documents, we retrieve the centroid vectors of each cluster. These centroids represent the average position of all document embeddings assigned to a particular cluster. For each cluster centroid, we find the nearest neighbors among the original Doc2Vec embeddings. These nearest neighbors are the documents that are closest to the centroid in the embedding space and can be considered as the main documents of that cluster. Thus we select for each cluster the 20 most representative documents  and we find a label for the topic on the basis of the documents titles. Note that the underlying meanings of the topics are subject to human interpretation. However, also this phase is automated by asking the topic to ChatGPT  (GPT-3.5) after having provided the list of 20 titles (see \cite{ebinezer20023nlp}).

Given the size of our sample, a reduction down to 30 topics can be considered a good compromise for a topic analysis. The selected number of topics strikes a good balance between ensuring a sufficient quantity of documents for each topic and maintaining the desired level of granularity. This approach allows us to extract meaningful insights from the data while avoiding an excessive division of content that could affect our ability to identify overarching trends and patterns. As shown in Figure \ref{fig:Topics}, most of the topics have an increasing trend (topic 28 seems to be the only exception). This is also motivated by the growth in the number of papers shown in Figure \ref{fig:Categories}. For each topic, the list of 20 titles is the input for the question we ask to the ChatGPT chatbot. The topics label (i.e. the ChatGPT reply to the question) together with the title of the most representative paper are reported in Table \ref{table:Topics}.  

It is interesting to observe that topics related to decentralized finance and blockchain technology (2) and stock price prediction with deep learning and news sentiment analysis (20) show a remarkable increase. This is also true for  health, policy, and social impact studies, represented by the topic 16, and diverse perspectives in education, innovation, and economic development (0). Both topics are oriented towards economics. These topics trends depend also by the introduction in 2014 of the q-fin.EC category within the arXiv quantitative finance papers. Classical quantitative finance subjects like portfolio optimization techniques and strategies (6), stochastic volatility modeling and option pricing (7), game theory and strategic decision-making (19)  high-order numerical methods for option pricing in finance (23) as well as, new themes appeared in the literature in the last years, like deep reinforcement learning in stock trading and portfolio management (4) and environmental and economic impacts of mobility technologies (25)  have attracted the interest of researchers in the analyzed period. The representativeness of topic 28 is limited, mainly because the number of papers in this cluster is low, and one could merge it with another cluster (i.e. 8). 

It is clear that some topics are more related to economics than to finance. This also depends on the presence of the q-fin.EC category. For articles in this category there is not always a flawless alignment with quantitative finance. 

To visualize the clusters, in Figure \ref{fig:TopicsClusters} the documents vectors are projected in the 2-dimensional space through the standard t-distributed stochastic neighbor embedding (t-SNE) algorithm. The larger the distance between topics, the more  distinct the papers in those topics are in the original high-dimensional space. Conversely, it is also possible to have a better view on how close are topics each other (e.g. 8 and 28). It appears that the most specialized or narrowly-focused topics tend to occupy peripheral positions, while themes that are more aligned with economics are positioned closer to the center.

\section{Extracting authors and journals}\label{sec:Authors}

In this section we extract the author surnames by means of the {\it spacy} package. More in details, starting from the raw data, we perform a named-entities recognition. Since this approach extract both first names and surnames, we remove all first names by checking if these names are included in a list of about 67,000 first names. It should be noted that the number of occurrence of the surname of the author in a paper strongly depends on the citation style. Surnames are always reported in the references, but they do not necessary appear in the main text of a paper. Additionally, even if the author-date style is widely used (i.e. the citation in the text consists of the authors name and year of publication), the surname of the first author appears more frequently (e.g. Bianchi is more probable than Tassinari, even if Bianchi and Tassinari are coauthors of the same papers, together with other coauthors).

The algorithm is able to find the names and surnames occurring in the text. These are included in the PERSON entity types. The first 100 authors by number of occurrences in the corpus are selected. In order to have additional information on these authors, we obtain topics, number of citations, h-index and i10-index from Google Scholar (see Table \ref{table:Authors}). It should be noted that not all authors are registered in Google Scholar, even if they made a significant contribution to the field (e.g. Markowitz) or there are authors with the same surname and belonging to the same research field (see also Figure 3 in \cite{sharma2023review}). This is the case for some researchers we find in our corpus (e.g. Zhou, Bayraktar and Chakrabarti). For these last authors is not simple to find a perfect match in Google Scholar even if their number of occurrences is generally high.\footnote{ For the reasons described above we exclude from Table \ref{table:Authors} the following researcher: Zhou, Markowitz, Peng, Jacod, Merton, Guo, Lo, Follmer, Yor, Almgren, Embrechts, Bayraktar, Artzner, Weber, Jarrow, Feng, Samuelson, Tang, Chakrabarti, Glasserman, Tsallis, Leung, Sato, Zariphopoulou, Kramkov, Karoui, Cizeau, Cao and Christensen.}

The algorithm is also able to find the most cited journals, included in the ORG entity types: Journal of Finance (4490 occurrences)$^*$, Mathematical Finance (3785), Journal of Financial Economics (3325)$^*$, Physica A (3137)$^*$, Quantitative Finance (3044), Econometrica (2473)$^*$, Journal of Econometrics (1971),  American Economic Review (1878), Insurance Mathematics and Economics (1667), Review of Financial Studies (2636)$^*$, Journal of Banking and Finance (1542)$^*$, Physical Review (1538), Journal of Economic Dynamics (1289), Energy (1267), Operations Research (1242), The Quarterly Journal of Economics (1238)$^*$, Journal of the American Statistical Association (1204), Management Science (1160), European Journal of Operational Research (1066), Quantum (1043), IEEE Transactions (996), Journal of Political Economy (990), Journal of Economic Theory (977), Energy Economics (946), International Journal of Theoretical and Applied Finance (946), SIAM Journal on Financial Mathematics (888), Science (865), Expert Systems with Applications (845), Applied Mathematical Finance (743), Finance and Stochastics (736),  Mathematics of Operations Research (652), PLoS (602), The Annals of Applied Probability (593), Stochastic Processes and their Applications (525), Energy Policy (520), International Journal of Forecasting (520), The European Physical Journal (514), Journal of Empirical Finance (510), Journal of Risk (509). We consider only journals with more than 500 occurrences and we exclude publishing houses. The journals with the asterisk symbol are identified with more than one name. It should be noted that some well-known journals are slightly below 500 occurrences. Furthermore, it is worth noting that papers with a strong mathematical focus tend to receive significantly fewer citations compared to papers that lean more towards economics or finance.

It is important to acknowledge that while the arXiv repository serves as a valuable resource for scholarly papers, it may not encompass the entirety of the quantitative finance research landscape. While the repository strives to be inclusive and comprehensive, there may be variations in the representation of scholars from different countries. Some scholars may have a relatively higher presence due to their active participation in submitting their research to arXiv. The platform content is reliant on authors voluntarily submitting their work, which introduces the possibility of selection bias. As a result, some authors and their contributions may not be represented. Therefore, our analysis and conclusions should be interpreted within the context of the available arXiv data, recognizing that there may be additional research and authors in the field of quantitative finance who have chosen alternative avenues for publishing their work. The same observation is true for the findings described in Section \ref{sec:Study}.

It is possible that influential scholars may not be as consistently represented or that, for various reasons, they have not regularly submitted their work to arXiv (see also \cite{metelko2023exploring}). In the study of \cite{sharma2023review}, focused on option pricing, some well-known authors are cited but they do not appear among the first 100 authors in our analysis. This discrepancy could be influenced by factors such as publication preferences, possible copyright issues, or institutional practices that may vary across different academic communities. 

As a final remark, in our view, researchers in quantitative finance should consider submitting their work to arXiv due to the potential benefits it offers (see also \cite{mishkin2020arxiving}). The delay between arXiv posting and journal publication, which can sometimes be more than a year, underscores the importance of submitting preprints to the repository. By doing so, researchers can help the community to understand research trends in their field more promptly, while also accelerating the dissemination of their own findings. This approach aligns with the findings of \cite{wang2020preprints}, who show that rapid and open dissemination through preprints helps scholarly and scientific communication, enhancing the reception of findings in the field.  In the meanwhile, as a possible alternative to gain a comprehensive understanding of the entire landscape, future studies may consider incorporating other reputable academic databases and journals to ensure a more holistic exploration of quantitative finance research and its authors. Network approaches would also help to identify cliques and highly connected groups of authors. Insights from network clusters could further improve the understanding of the textual data investigated in this work.

\section{Conclusions}\label{sec:Conclusions}

In this study, we explore the field of quantitative finance through an analysis of papers in the arXiv repository. Our objectives are twofold: first, we investigate the evolution of topics over time, second, we identify prominent authors and journals in this domain. By employing data mining techniques, we achieve these goals without reading the papers individually. 

The preprocessing phase, when needed, ensures the suitability of the data for subsequent analyses. Topic modeling helps in gaining insights and understanding the main themes and trends within our large dataset. By applying topic modeling algorithms, we identify the best performer and examine the temporal evolution of quantitative finance topics. This analysis reveals the changing research trends and highlights the emergence and decline of various topics over time. 

Furthermore, we conduct an entity extraction process to identify influential authors and journals in the field. Through quantifying authors and journals occurrences, we shed light on the researchers who have made notable contributions to quantitative finance.

Our study demonstrates the power of data mining techniques in uncovering insights from a large-scale preprint repository. Our work not only showcases the power of data mining but also highlights the continued growth and dynamism of quantitative finance as a discipline. The techniques explored in this work can assist researchers in exploring and identifying novel research topics, discovering connections between different research areas, and staying up-to-date with the latest developments in the field. Furthermore, our methodology may serve as a roadmap for future studies on broader datasets or in other scientific domains  utilizing text mining techniques. Although scaling to a larger number of papers may pose challenges, our approach provides valuable insights.

Finally, we believe that quantitative finance researchers should consider sharing their work on arXiv to potentially accelerate the dissemination and impact of their findings and to enhance the community understanding of research trends.

\newpage


\begin{landscape}
\sffamily{%
\tiny
\begin{longtable}{cll}
\toprule
number &  label  & title of the most representative paper \\
\midrule
\endhead
0 & diverse perspectives in education, innovation, and economic development & Perspectives in public and university sector co-operation in the change
  of higher education model in Hungary, in light of China's experience\\
1 & modeling financial market dynamics & Comment on: Thermal model for adaptive competition in a market \\
2 & decentralized finance and blockchain technology & Understanding the maker protocol\\
3 & correlation analysis in financial markets and networks & Random matrix theory and cross-correlations in global financial indices
  and local stock market indices\\
4 & deep reinforcement learning in stock trading and portfolio management & Practical deep reinforcement learning approach for stock trading
optimal market making by reinforcement learning\\
5 & optimal trading and portfolio liquidation strategies in financial markets & An FBSDE approach to market impact games with stochastic parameters\\
6 & portfolio optimization techniques and strategies & Seven sins in portfolio optimization\\
7 & stochastic volatility modeling and option pricing & On the uniqueness of classical solutions of Cauchy problems\\
8 & asset pricing, investment, and arbitrage in financial markets & Characterization of arbitrage-free markets \\
9 & network analysis of financial contagion and systemic risk & Clearing algorithms and network centrality\\
10 & counterparty risk and valuation adjustments in financial derivatives & Collateral margining in arbitrage-free counterparty valuation adjustment
  including re-hypotecation and netting \\
11 & quantum models in finance and option pricing & Sornette-Ide model for markets: Trader expectations as imaginary part\\
12 & valuation and risk management in annuity and insurance products & A policyholder's utility indifference valuation model for the guaranteed
  annuity option\\
13 & optimal dividend strategies in stochastic control and risk management & Optimal dividends problem with a terminal value for spectrally positive
  Levy processes\\
14 & risk measures and utility maximization under model uncertainty & On the C-property and $w^*$-representations of risk measures\\
15 & economic complexity, networks, and trade patterns & Economic complexity and growth: Can value-added exports better explain the link?\\
16 & health, policy, and social impact studies & Ramadan and infants health outcomes \\
17 & statistical analysis of financial markets and volatility & Volatility distribution in the S\&P500 Stock Index\\
18 & renewable energy economics and electricity market dynamics & On wholesale electricity prices and market values in a carbon-neutral energy system\\
19 & game theory and strategic decision-making & Simultaneous auctions for complementary goods \\
20 & stock price prediction with deep learning and news sentiment analysis & Stock Prediction: a method based on extraction of news features and
  recurrent neural networks \\
21 & kinetic wealth exchange models in economics & Gibbs versus non-Gibbs distributions in money dynamics\\
22 & market order flow and price impact & Order flow and price formation \\
23 & high-order numerical methods for option pricing in finance & High-order compact finite difference scheme for option pricing in
  stochastic volatility with contemporaneous jump models\\
24 & optimal investment and consumption in financial models with constraints & Recursive utility optimization with concave coefficients\\
25 & environmental and economic impacts of mobility technologies & A review on energy, environmental, and sustainability implications of
  connected and automated vehicles\\
26 & advanced risk measures in financial modeling& Generating unfavourable VaR scenarios with patchwork copulas\\
27 & Bayesian models for financial tail risk forecasting& A semi-parametric realized joint value-at-risk and expected shortfall
  regression framework\\
28 & pricing and modeling options in stochastic volatility models with jumps & Semi-analytical pricing of barrier options in the time-dependent Heston
  model\\
29  & economic growth and market dynamics & Uncovering volatility dynamics in daily REIT returns\\

\bottomrule
\caption{\tiny For each topic we report the label extracted from ChatGPT and the title of the most representative paper.}
\label{table:Topics}
\end{longtable}}
\end{landscape}

\begin{landscape}
\sffamily{%
\tiny
\begin{longtable}{lrlllll}
\toprule
author\_id &  occurences &name &topics & citations & h-index & i10-index \\
\midrule
\endhead
  mG07\_6k &      4505.0 &        Ioannis Karatzas &                                                              ['stochastic analysis', 'stochastic control', 'mathematical finance'] &     35724 &      60 &       127 \\
  58amEmw &      4441.0 &  Jean-Philippe Bouchaud &                   ['statistical mechanics', 'disordered systems', 'random matrices', 'quantitative finance', 'agent based models'] &     49794 &     105 &       351 \\
  ahLm1v0 &      3256.0 &    Walter Schachermayer &                                                                                                                                 [] &     15891 &      56 &       124 \\
  HGsSmMA &      3135.0 &         Didier Sornette &                                                                          ['cooperation', 'organization', 'patterns', 'prediction'] &     53371 &     112 &       587 \\
  CqFCQVE &      2794.0 &        Alexander Schied &                                                             ['probability theory', 'stochastic processes', 'mathematical finance'] &     16780 &      35 &        71 \\
  2QOp9\_M &      2457.0 &              Peter Carr &                             ['financial engineering', 'quantitative finance', 'mathematical finance', 'derivatives', 'volatility'] &     24060 &      62 &       117 \\
  mVF1X\_U &      2371.0 &          Freddy Delbaen &                                                                                                         ['mathematik', 'ökonomie'] &     25557 &      50 &        90 \\
  ElAtiUs &      2258.0 &          Darrell Duffie &                                                                                                     ['finance', 'central banking'] &     58442 &      88 &       142 \\
  fq7BQos &      2042.0 &          Dilip B. Madan &                                                                             ['mathematical finance', 'general equilibrium theory'] &     25131 &      55 &       156 \\
  8abFiFM &      1774.0 &            Robert Engle &                                                                                                       ['finance and econometrics'] &    189678 &     118 &       229 \\
  GU9HgNA &      1743.0 &             Quanquan Gu &     ['statistical machine learning', 'nonconvex optimization', 'deep learning theory', 'reinforcement learning', 'ai for science'] &     14915 &      56 &       161 \\
  Q7N-rCk &      1672.0 & Rosario Nunzio Mantegna &                           ['econophysics', 'statistical physics', 'complex systems', 'financial markets', 'information filtering'] &     28269 &      67 &       139 \\
  QsYYhSE &      1629.0 &          Søren Johansen &                                                                                         ['matamatical statistics', 'econometrics'] &     97728 &      65 &       132 \\
  vZA2pjw &      1627.0 &    Benoît B. Mandelbrot &                                                   ['mathematics', 'fractals', 'economics', 'information theory', 'fluid dynamics'] &    142895 &      96 &       319 \\
  Lf1kf1Q &      1505.0 &            Mario Coccia &          ['evolution of technology', 'scientific change', 'social dynamics', 'complex adaptive systems', 'environment \& covid-19'] &     19376 &     106 &       228 \\
  9HXRjPk &      1444.0 &         Damir Filipovic &                                                                           ['quantitative finance', 'quantitative risk management'] &      6910 &      39 &        73 \\
  6\_INHZI &      1389.0 &          Fabrizio Lillo &                                                                  ['quantitative finance', 'statistical mechanics', 'data science'] &     10900 &      51 &       121 \\
  mGpnlA8 &      1218.0 &             Touzi Nizar &                                                              ['stochastic control', 'mathematical finance', 'monte carlo methods'] &     11831 &      56 &       120 \\
  rp-3Yoo &      1187.0 &          Barry Williams &                                                                     ['banks and banking', 'bank risk', 'multinational', 'banking'] &      2342 &      17 &        23 \\
  MZNxzRY &      1161.0 &              Huyên Pham &                                                          ['mathematical finance', 'stochastic control', 'numerical probabilities'] &      9967 &      54 &       135 \\
  3HhvEUc &      1147.0 &            Yuri Kabanov &                                                                                            ['mathematical finance', 'mathematics'] &      6297 &      38 &        77 \\
  -YEPo1E &      1143.0 &         Wing-Keung Wong &                            ['financial economics', 'econometrics', 'investment theory', 'risk management', 'operational research'] &     14440 &      65 &       274 \\
  GyPrRgc &      1138.0 &        Swarn Chatterjee &                       ['financial planning', 'wealth management', 'financial literacy', 'household finance', 'behavioral finance'] &      2719 &      28 &        54 \\
  RZid9X8 &      1075.0 &        Guido Caldarelli &                                                    ['network theory', 'network science', 'statistical physics', 'complex systems'] &     24165 &      71 &       191 \\
  ImhakoA &      1075.0 &         Daniel Kahneman &                                                                                                                                 [] &    519507 &     158 &       369 \\
  zO\_tShM &      1050.0 &         Marek Rutkowski &                                                                                   ['mathematical finance', 'stochastic processes'] &      7559 &      30 &        67 \\
  7NJ7Ax8 &      1039.0 &       Patrick Cheridito &                                                                                                                                 [] &      5400 &      34 &        59 \\
  nyfza90 &      1019.0 &          Volker Schmidt &   ['virtual materials testing', 'statistical learning', 'image analysis', 'spatial stochastic modeling', 'monte carlo simulation'] &     11896 &      52 &       230 \\
  x4vtSxI &      1017.0 &            Rene Carmona &                                                          ['stochastic analysis', 'financial mathematics', 'financial engineering'] &     17474 &      59 &       139 \\
  kukA0Lc &       999.0 &           Yoshua Bengio &                                                                   ['machine learning', 'deep learning', 'artificial intelligence'] &    656874 &     222 &       763 \\
  vQ0\_nz8 &       989.0 &          Emmanuel Bacry &            ['self-similarity', 'multifractal', 'stochastic modeling', 'statistical finance', 'financial time-series modelization'] &     11937 &      47 &        69 \\
  1XwLUrc &       980.0 &            Jim Gatheral &                                                            ['volatility modeling', 'market microstructure', 'algorithmic trading'] &      6126 &      30 &        42 \\
  3HwRbiQ &       955.0 &         Jerome Friedman &                                                                                                                                 [] &    283058 &      95 &       197 \\
  e2Xowj0 &       900.0 &           Neil Shephard &                                                   ['econometrics', 'economics', 'statistics', 'financial econometrics', 'finance'] &     42035 &      69 &       140 \\
  pEnxwCM &       887.0 &     Victor M. Yakovenko &                                                                                        ['condensed matter theory', 'econophysics'] &      8891 &      44 &       101 \\
  a11vssU &       845.0 &   Constantinos Kardaras &                                                                     ['stochastic analysis', 'probability', 'mathematical finance'] &      1557 &      20 &        31 \\
  79htA7g &       838.0 &          Bent Flyvbjerg &                                                       ['project management', 'management', 'infrastructure', 'planning', 'cities'] &     73264 &      70 &       152 \\
  zH1qBSo &       834.0 &         Albert Shiryaev &                                                                                                             ['probability theory'] &     35521 &      59 &       163 \\
  QVb4LGI &       815.0 &            Andrey Itkin &                       ['mathematical finance', 'computational finance', 'derivatives', 'quantitative finance', 'machine learning'] &       709 &      14 &        19 \\
  Zuhod6s &       813.0 &               Yong Deng &                               ['uncertainty', 'deng entropy', 'information volume', 'random permutation set', 'chaos and fractal'] &     23189 &      81 &       335 \\
  bWlZ3-Y &       810.0 &           Eric Jacquier &                                                                                                                                 [] &      4458 &      19 &        25 \\
  GKthQJQ &       804.0 &           Peter K. Friz &                                                                    ['rough path theory', 'stochastic analysis', 'pdes', 'finance'] &      5678 &      39 &        82 \\
  2qTa\_4U &       794.0 &         Francis Diebold &                                                                         ['economics', 'econometrics', 'time series', 'statistics'] &     76159 &      97 &       175 \\
  utY1nTo &       794.0 &          Matteo Marsili & ['statistical mechanics', 'stochastic processes', 'collective phenomena in socio-economic systems', 'networks', 'complex systems'] &     10093 &      50 &       139 \\
  ZpG\_cJw &       783.0 &       Robert Tibshirani &                                                                                 ['statistics', 'data science', 'machine learning'] &    460493 &     172 &       525 \\
  65wdZxA &       780.0 &           Damiano Brigo &        ['probability', 'mathematical finance', 'stochastic analysis', 'signal processing', 'differential geometry and statistics'] &      9663 &      42 &       114 \\
  bxJe87s &       780.0 &         Marco Frittelli &                                                                   ['financial mathematics', 'mathematical finance', 'probability'] &      3795 &      24 &        33 \\
  aVju7cI &       771.0 &       Monique Jeanblanc &                                                                                                      ['mathématiques financières'] &     10256 &      51 &       110 \\
  -iOn6uI &       769.0 &        Aurélien Alfonsi &                                                                                                                                 [] &      2731 &      22 &        30 \\
  P\_LECrk &       750.0 &      Tomasz R. Bielecki &                       ['mathematical finance', 'stochastic processes', 'stochastic control', 'stochastic analysis', 'probability'] &      6901 &      39 &        89 \\
  5sQ0Fag &       729.0 &              Ajit Singh &                                                                                                                                 [] &     21592 &      60 &       360 \\
  6quAJUE &       706.0 &         Josef Teichmann &                                                          ['mathematical finance', 'machine learning in finance', 'rough analysis'] &      3019 &      30 &        67 \\
  58amEmw &       705.0 &  Jean-Philippe Bouchaud &                   ['statistical mechanics', 'disordered systems', 'random matrices', 'quantitative finance', 'agent based models'] &     49794 &     105 &       351 \\
  JicYPdA &       691.0 &         Geoffrey Hinton &                             ['machine learning', 'psychology', 'artificial intelligence', 'cognitive science', 'computer science'] &    687453 &     180 &       436 \\
  i2MC67A &       679.0 &               J.F. Muzy &                                                                            ['multifractal analysis', 'econophysics', 'turbulence'] &     13417 &      55 &        83 \\
  K9yGky8 &       678.0 &       Andreas Kyprianou &                                                                                      ['probability theory', 'applied mathematics'] &      7946 &      44 &        99 \\
  aCSds20 &       670.0 &           Xavier Gabaix &                                                                                                           ['economics', 'finance'] &     30926 &      56 &        75 \\
  G-WPCrM &       667.0 &      Diego Garlaschelli &                                                          ['network theory', 'econophysics', 'sociophysics', 'statistical physics'] &      7394 &      41 &        73 \\
  YTCnA4E &       664.0 &        Eduardo Schwartz &                                                                                                                        ['finance'] &     44652 &      81 &       141 \\
  A0ISJPU &       664.0 &           Steven Shreve &                                                                                           ['probability', 'financial mathematics'] &     35605 &      42 &        70 \\
  fFFOHec &       660.0 &        Alexander McNeil &                                                                                                                                 [] &     24473 &      41 &        60 \\
  dYwbc9s &       659.0 &            Guido Imbens &                                                                                               ['causal inference', 'econometrics'] &     90765 &      95 &       169 \\
  OQK4DDY &       657.0 &           Peter Forsyth &                                                    ['scientific computing', 'computational finance', 'numerical solution of pdes'] &     10617 &      58 &       143 \\
  c1wQ9\_k &       655.0 &            Daojian Zeng &                                                                                                    ['natural language processing'] &      4877 &      13 &        17 \\
  Vs7kOf4 &       645.0 &             Marcel Nutz &                                                                       ['optimal transport', 'mathematical finance', 'game theory'] &      2339 &      30 &        43 \\
  vjc1kF0 &       640.0 &       Francesca Biagini &                                                      ['financial and insurance mathematics', 'stochastic calculus', 'probability'] &      2598 &      20 &        42 \\
  zGJKZpk &       629.0 &       Marianne Bertrand &                                                                                                                                 [] &     64693 &      66 &       119 \\
  nEfnJZM &       628.0 &          Vadim Linetsky &                                                                                    ['mathematical finance', 'financial economics'] &      5260 &      39 &        63 \\
  Bekg2Qo &       621.0 &            Joel Shapiro &            ['financial intermediation', 'regulation of financial institutions', 'corporate governance', 'industrial organization'] &      3449 &      15 &        16 \\
  KDhGvNQ &       611.0 &          Johanna Ziegel &                              ['statistical forecasting', 'risk measures', 'postitive definite functions', 'stereology', 'copulas'] &      2088 &      19 &        31 \\
  r5PHkCs &       610.0 &             Thomas Guhr &                                                                                                            ['theoretical physics'] &      8395 &      39 &       104 \\
\bottomrule
\caption{\tiny Number of authors occurrences and Google Scholar metrics as of May 2023.}
\label{table:Authors}
\end{longtable}}
\end{landscape}

\end{document}